\documentclass[preprint,A4paper,superscriptaddress,pre]{revtex4-1}
\usepackage{amsmath,amsfonts,amssymb}
\usepackage{graphicx}
\usepackage{epstopdf}
\usepackage{hyperref}

\newcommand{\sig}[1]{\mathrm{Sign}#1}

\def\ud{\mathrm{d}}
\def\B{\beta}

\def\vecxmu{\vec{x}^\mu}
\def\vecj{\vec{J}}
\def\vecJ{\vec{J}}

\def\vecJcero{\vec{J^0}}

\def\hatq{\hat{q}}
\def\hatr{\hat{r}}
\def\hatQ{\hat{Q}}

\def\intrho{\int \ud J^0 \rho(J^0)}
\def\DNx{\mathrm{D}x\:}

\def\neffcero{n_{\mbox{\tiny eff}}^{0}}

\newcommand{\DN}[1]{\mathrm{D}#1\:}

\newcommand{\norm}[1]{\| #1 \|_p}

\def\H{\mathcal{H}}

\begin{document}
 \title{ Stability of the replica symmetric solution in diluted perceptron learning}

\author{Alejandro Lage-Castellanos}
\email[Corresponding e-mail: ]{lage at fisica+uh+cu}

\author{Gretel Quintero Angulo}
\affiliation{Department of Theoretical Physics, Physics  Faculty, University of Havana, La Habana, CP 10400, Cuba}
\affiliation{``Henri-Poincar\'e'' Group of Complex Systems and Department of Theoretical Physics, Physics Faculty, University of Havana, La Habana, Cuba.}

\author{Andrea Pagnani}
\affiliation{Human Genetic Foundation  (HuGeF-Torino), Via Nizza 52, I-10122, Torino, Italy}

\date{\today}


\begin{abstract}

We study the role played by the dilution in the average behavior of a
perceptron model with continuous coupling with the replica method.  We
analyze the stability of the replica symmetric solution as a function
of the dilution field for the generalization and memorization
problems.  Thanks to a Gardner like stability analysis we show that
at any fixed ratio $\alpha$ between the number of patterns $M$ and the
dimension $N$ of the perceptron ($\alpha=M/N$), there exists a
critical dilution field $h_c$ above which the replica symmetric ansatz
becomes unstable.

\end{abstract}



\maketitle

\section*{Introduction}

Neural networks \cite{statistical_learning_shibirani} have become among the
most studied and successful models in the field of
artificial intelligence. In spite of more than 50 years of research,
the field is still thriving
\cite{learning_zecchina05,andrea,andrea2}. In the last years, thanks
to the advances of the new-generation high-throughput technologies in
molecular biology, the field has experienced a renewed interest with
problems coming from the analysis of high dimensional data in
biology, where sparsity in the underlying model is a general feature
\cite{kabashima,kabashima2,mezard_compress,LASSO,andrea,andrea2}.

In this paper we address the issue of the stability of the replica
symmetric solution presented in \cite{lagesparse09} which has
important implications for the implementation of many algorithmic
strategies. This paper extends thus the analysis presented in
\cite{lagesparse09} and generalizes the stability results of
\cite{hungaros} to the external dilution case.

A standard problem in artificial intelligence is that of
generalization \cite{vallet,hungaros}. A set of $M$ patterns $\vecxmu$
is given together with a binary output variable $y^\mu$ for each of
them ($\mu \in \{1,2,\dots,M\}$). A pattern $\vec x$ is an $N$
dimensional vector and we are interested in learning the hidden
relation among its components and the classification value $y^\mu$
related to pattern $\mu$. In the simplest setup, a linear perceptron
tries to encode such relation in a vector $\vecJ$ (called {\it
  student}) such that the binary classification $y^\mu = \pm 1$ is
reproduced by
\[ y^\mu = \sig(\vecj\cdot\vecxmu)
\]
We will consider that the classification is actually generated by an
unknown {\it teacher} $\vecJcero$
\begin{equation}
  y^\mu = \sig(\vecJcero\cdot\vecxmu + \eta^\mu) \label{eq:noise}
\end{equation}
and therefore, the aim of the student $\vecJ$ is to be as close as
possible to the teacher $\vecJcero$.  A Gaussian noise $\eta^\mu \sim N(0,\gamma^2)$ is added to the classification function to account for experimental noise in the data. Two interesting limits are studied: no noise $\gamma = 0$, and random classification $\gamma = \infty$.


As a first step to analyze the problem, we define an energy function
counting the number of patterns that are wrongly classified by the
student $\vecJ$:
\begin{equation}
E(\vecj)=\sum_\mu^M \Theta(-y^\mu \vecj\cdot\vecxmu ) .\label{eq:E}
\end{equation}
In the following we will consider only the case
$J\in {\cal R}^N$. Since the energy expression depend only on the angle of $\vecJ$
and not on his length, {\em i.e.} $E(c \vecJ) = E(\vecJ)$ for all
$c>0$, we restrict ourself to the surface of a sphere $\sum_{i=1}^N J_i^2
= N$.

Standard statistical mechanics method
\cite{statistical_mechanics_of_learning, gardner87,
  gardner88,andrea,kabashima2} have been largely used to study the
thermodynamic properties of this problem. In a recent work
\cite{lagesparse09} a slightly different point of view has been
analyzed: which is the $\vecJ$ that minimizes the energy function with
the largest possible number of coordinates equal to zero or, in other
words, which is the sparsest possible $\vecJ$ compatible with a
correct classification? To address this issue one needs to study the
performance of perceptron in presence of an external dilution field
$h$ coupled to the classification vector $\vecJ$. To this end one can
consider the following Hamiltonian:
\begin{equation}
\B\H(\vecj)=\B E(\vecJ) + h \norm{\vecJ} \label{eq:Ham}
\end{equation}
where the dilution field $h$, in the last term, acts as a chemical
potential on $\vecJ$. Two interesting cases were studied at the
replica symmetric level: the $L_1$ norm ($\| \vecJ \|_1 = \sum_{i=1}^N
|J_i|$) and the $L_0$ norm ($\|\vecJ \|_0 = \lim_{p\to 0} \sum_{i=1}^N
|J_i|^p= \sum_{i=1}^N(1-\delta_{J_i})$).

Of course this problem has practical interest in the case one knows
{\em a priori} when the teacher is actually sparse. In many real life
problems \cite{andrea, andrea2,kabashima,mezard_compress} the
multidimensional patterns $\vec x$ contain irrelevant information, in
the sense that only few of the components are actually considered in
the classification process. We will consider that only a fraction
$\neffcero$ of the teacher's components are non zero. Therefore each
teacher's component is extracted independently from the distribution
\begin{equation}
 \rho(J_0) = (1-\neffcero) \delta_{J_0} + \neffcero \rho'(J_0). \label{eq:rhoJcero}
\end{equation}
It has been shown \cite{LASSO,kabashima} that dilution with $L_p$
norm, with $p\leq 1$, forces a fraction of the components of $\vecJ$
to be exactly zero, and therefore, allows a selection of the
components that actively participate in the classification process.

The cost function $\H(\vecJ)$ raises severe computational problems due
to the scale invariance of the energy which makes the optimization
problem non-convex in general. At odd with problems like compressed
sensing \cite{Donoho2006}, for which optimization methods of order
$O(N^3)$ can be used for the $L_1$ norm \cite{kabashima} whereas the
most effective $L_0$ norm \cite{kabashima} makes the minimization an
NP-hard optimization problem \cite{andrea, kabashima}, in our
perceptron-like case we are not aware of any {\em ad-hoc} numerical
technique for finding efficiently the minimum of the cost function in
Eq.~(\ref{eq:Ham}), apart from Monte Carlo based optimization methods.

From a theoretical perspective, we are interested in the average case
properties of the structure of the solution space in either cases $L_1$
and $L_0$.  The paper is organized in the following way: the next
section (\ref{sec:rs}) fixes the notation and sketches the replica
symmetric results presented in \cite{lagesparse09}, in section
\ref{sec:gardner} we study the stability of this replica symmetric
solution, in section \ref{sec:cases} we show that at $h=0$ the replica
symmetric solution is always stable, while at $h\to\infty$ is always
unstable, and therefore a critical value of the dilution field marks
the threshold between two phases, computed in section
\ref{sec:phase}. Finally, our results are summarized in section
\ref{sec:conclusions}, and the main results of \cite{lagesparse09} are
re-evaluated in the present context.

\section{Replica symmetric solution to diluted perceptron}
\label{sec:rs}

Using the replica method \cite{MPV,statistical_mechanics_of_learning}
one can compute the average free energy as the limit $ -\B
\overline{f}= \lim_{n\to 0}$ $\lim_{N\to \infty}
\log\overline{ Z^n}/nN$. The replica symmetric (RS) ansatz assumes that
any two replicas of the system have the same overlap, and therefore
the corresponding overlap matrix and its Fourier conjugate assume the
following structure:
{\small
\begin{equation}
Q_{a,b}=\left(\begin{array}{ccccc}
t & & & &\\
r & 1 & & &\\
r & q & 1 & &\\
. & . & . & 1 &\\
r & q & . & q & 1
\end{array}\right) \quad -i \hat{Q}_{a,b}=\left(\begin{array}{ccccc}
\lambda^0 & & & &\\
\hatr & \lambda & & &\\
\hatr & \hatq & \lambda & &\\
.     &   .   &    .    & \lambda &\\
\hatr & \hatq & . & \hatq & \lambda
\end{array}\right)  \label{eq:replica symmetricoverlap}
\end{equation}
}

These are symmetric $(n+1) \times (n+1)$ matrices where the first
column (row) contains the teacher's parameters. For instance, $t$
is the variance of the teacher and is related to the distribution
(\ref{eq:rhoJcero}), and is not a variational parameter. On the other
hand, $r = \langle \frac 1 N \vecJ \cdot \vecJcero \rangle$ is the average
overlap between the teacher and the student, and $q = \langle \frac 1
N \vecJ^a \cdot \vecJ^b \rangle$ is the average overlap between two
replicas (see \cite{lagesparse09} for details).

The correct value for the free energy is obtained by minimizing the variational RS free energy
\begin{equation} -\B \overline{f}=-r\hatr+\frac{1}{2}q \hatq -\lambda + G_J+\alpha\:  G_X \nonumber
\end{equation}
with respect to the variational parameters $(q,r,\hatq,\hatr,\lambda)$, where
{\small
\begin{eqnarray}
G_J & =&\int \DNx \intrho \log\int \ud J e^{-(\frac{\hatq}{2}-\lambda)J^2 -h \norm{J} +(\hatr J^0- \sqrt{\hatq} x) J} \label{eq:Gs} \\
 G_X & = & 2 \int \DNx H\left(\frac{ x r}{\sqrt{q \gamma^2+q t-r^2)} }\right) \log\left((e^{-\B}-1) H(-\sqrt{\frac{q}{1-q}}x)+1 \right) \nonumber
\end{eqnarray}
}
In these equations, and from now on, $\DNx =  \ud x\: \exp(-x^2/2)/\sqrt{2 \pi}$ is a Gaussian measure in $x$, and the function $H(y) = \displaystyle \int_y^\infty \DNx$.

A key role is played by $\alpha = \frac{M}{N}$, the ratio between the
number of patterns $M$ and the space dimensionality $N$, that measures
the amount of information we have and, together with $h$, is the
fundamental parameter controlling the quality of the
generalization. We assume that in the limit $N \rightarrow \infty$, $\alpha$ remains finite.




\section{Stability {\it a la} Gardner}
\label{sec:gardner}

The main goal of this work is to analyze the stability of the replica
symmetric solution. Other structures for the overlap matrices
(\ref{eq:replica symmetricoverlap}) are indeed possible. In analogy
with the organization of the thermodynamic states in the low temperature
phase of mean field spin glass-like models \cite{MPV}, we expect
that the space of solutions of our model could be described by a
hierarchical (ultrametric) organization of replicas, or, in other terms,
that our system might undergo a spontaneous replica symmetry breaking
(RSB). From a purely geometric point of view this process is related
to the fragmentation of the space of zero energy configurations into
unconnected regions, or equivalently, the fragmentation of the Gibbs
measure into many pure states:
\begin{figure}[!htb]
\includegraphics[scale=0.5, angle=0]{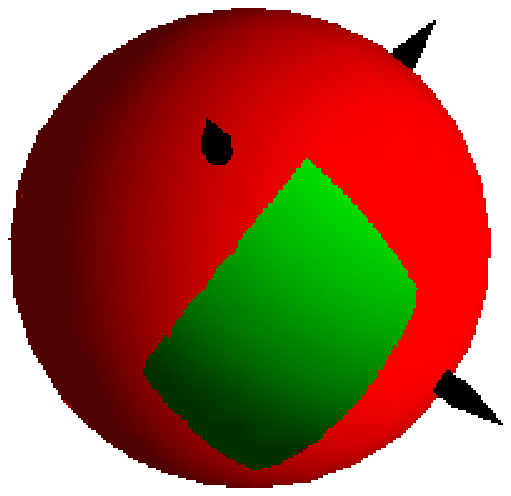}
\hspace{1cm}
\includegraphics[scale=0.5, angle=0]{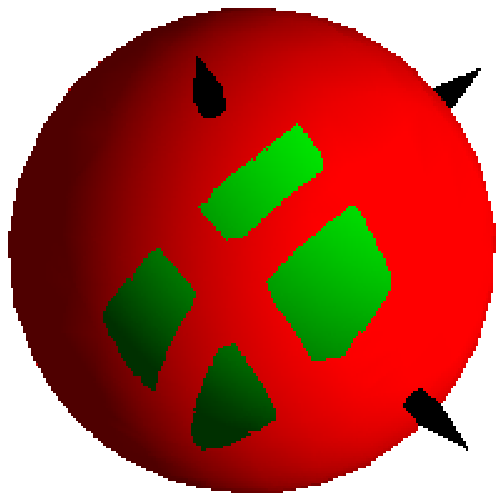}
\caption{Representation of a connected (left) and a disconnected
  (right) solution space on the sphere. Connected phase space is
  typical of replica symmetric solutions, while the breaking in the
  symmetry of the replicas implies the clusterization of the space.}
\label{fig:spheres}
\end{figure}

We will now apply Gardner's method for the stability analysis of
replica symmetric solution in \cite{gardner87}, studying the
eigenvalues of the Hessian matrix:

{\small\begin{equation}
\|\partial^2(\B n \overline{f_n})\| = \left(\begin{array}{ccccccc}
\|\frac{\partial^2(\B n \overline{f_n}) }{\partial Q_{ab} \partial Q_{cd}}\|&&&\|\frac{\partial^2(\B n \overline{f_n}) }{\partial Q_{ab} \partial (-\imath \hatQ_{cd})}\|\\
\|\frac{\partial^2(\B n \overline{f_n}) }{\partial Q_{ab} \partial (-\imath \hatQ_{cd})}\|&&&\|\frac{\partial^2(\B n \overline{f_n}) }{\partial (-\imath \hatQ_{ab}) \partial (-\imath \hatQ_{cd})}\|
\end{array}\right) \label{eq:hessian}
\end{equation}}

evaluated at the RS fixed point. Details of the calculation are given
in the Appendix \ref{app:details}.

The eigenvalues of the $n (n+1)\times n (n+1)$ Hessian matrix are
simply related to the spectrum of the four sub-matrices. Among the eigenvectors
only those with eigenvalues $\delta_1$ and
$\delta_2$, signal an instability of the  replica symmetric solution:

\begin{equation}
\delta_1 \delta_2 = \gamma_1 \gamma_2 - 1 \label{eq:eigenvalue_product}
\end{equation}
where $\gamma_1$ and $\gamma_2$ are eigenvalues of the inner matrices
$\|\frac{\partial^2(\B n \overline{f_n}) }{\partial Q_{ab} \partial
  Q_{cd}}\|$ and $\|\frac{\partial^2(\B n \overline{f_n}) }{\partial
  (-\imath \hatQ_{ab}) \partial (-\imath \hatQ_{cd})}\|$, given by
\begin{eqnarray}
\gamma_1 &=& - \frac{2 \alpha}{(1-q)^2} \int \DN{y} H\left(\frac{y r}{\sqrt{q \gamma^2 + q t - r^2}}\right) \left\{ \left(\frac{e^{-\frac{h_0^2}{2}}}{\sqrt{2 \pi} (H(h_0)-1)}\right)^2 - \frac{h_0 e^{-\frac{h_0^2}{2}}}{\sqrt{2 \pi} (H(h_0)-1)}\right\}^2 \nonumber \\
\gamma_2 &=& - \int \DN{z} \intrho \left\{\frac{\int \ud J J^2 e^{H_J}}{\int \ud J e^{H_J}}-\left(\frac{\int \ud J J e^{H_J}}{\int \ud J e^{H_J}}\right)^2\right\}^2.
\end{eqnarray}
where $H_J = -(\frac{\hatq}{2}-\lambda) J^2 + (\hatr J_0 - \sqrt{\hatq} x)J - h|J|^p$.

The sign of the product $\delta_1 \delta_2$ determines the stability
of the extremal point: when $\delta_1 \delta_2 < 0$ the point is
stable, and unstable otherwise.

\section{Extreme cases: $h=0$ and $h\to \infty$}
\label{sec:cases}

In the zero temperature case, the Gibbs measure concentrates over the
states of lower energy, {\em i.e.} over those vectors $\vecJ$ that
correctly classify most patterns. In the absence of dilution ($h=0$),
the measure is uniform over these states, and the free energy is thus
a measure of their entropy (we are working at zero temperature). On
the other extreme, we have the high dilution limit $h\to\infty$ that
we studied in \cite{lagesparse09} at replica symmetric level. Next we
check the stability of the replica symmetric solution for both the
Memorization and Generalization problems in these two extreme
situations. As we will see, the replica symmetric solution is stable
for $h=0$ and unstable for $h\to\infty$, and therefore, there is a
critical value $h_c(\alpha)$ dividing the RS phase from the non-RS
one.

\subsection{Stability of non diluted perceptron $h=0$}

In the absence of dilution ($h=0$) we still have two cases:
memorization ($\gamma\to\infty$) and generalization ($\gamma =0$).  In
the high noise limit $\gamma \rightarrow \infty$ (see
eq. (\ref{eq:noise})) there is no rule to infer because the
experiments are randomly classified, and the perceptron tries to
memorizes the output variable $y^\mu$. In the $\gamma =0$ case, the
patterns are classified according to the teacher and this rule can be
generalized.

\begin{figure}[!htb]
\includegraphics[scale=0.3,
  angle=270]{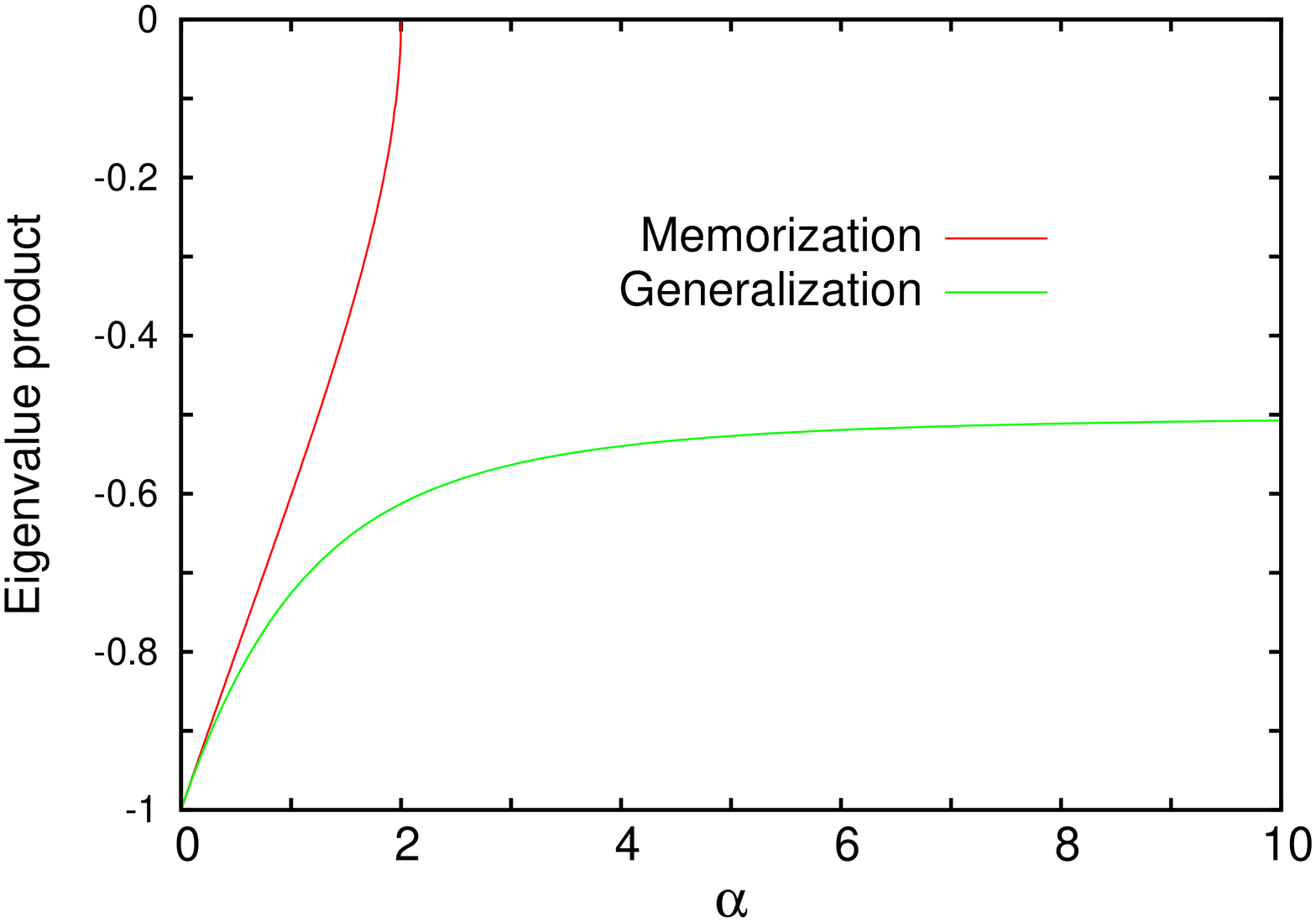}
\includegraphics[scale=0.3,
  angle=270]{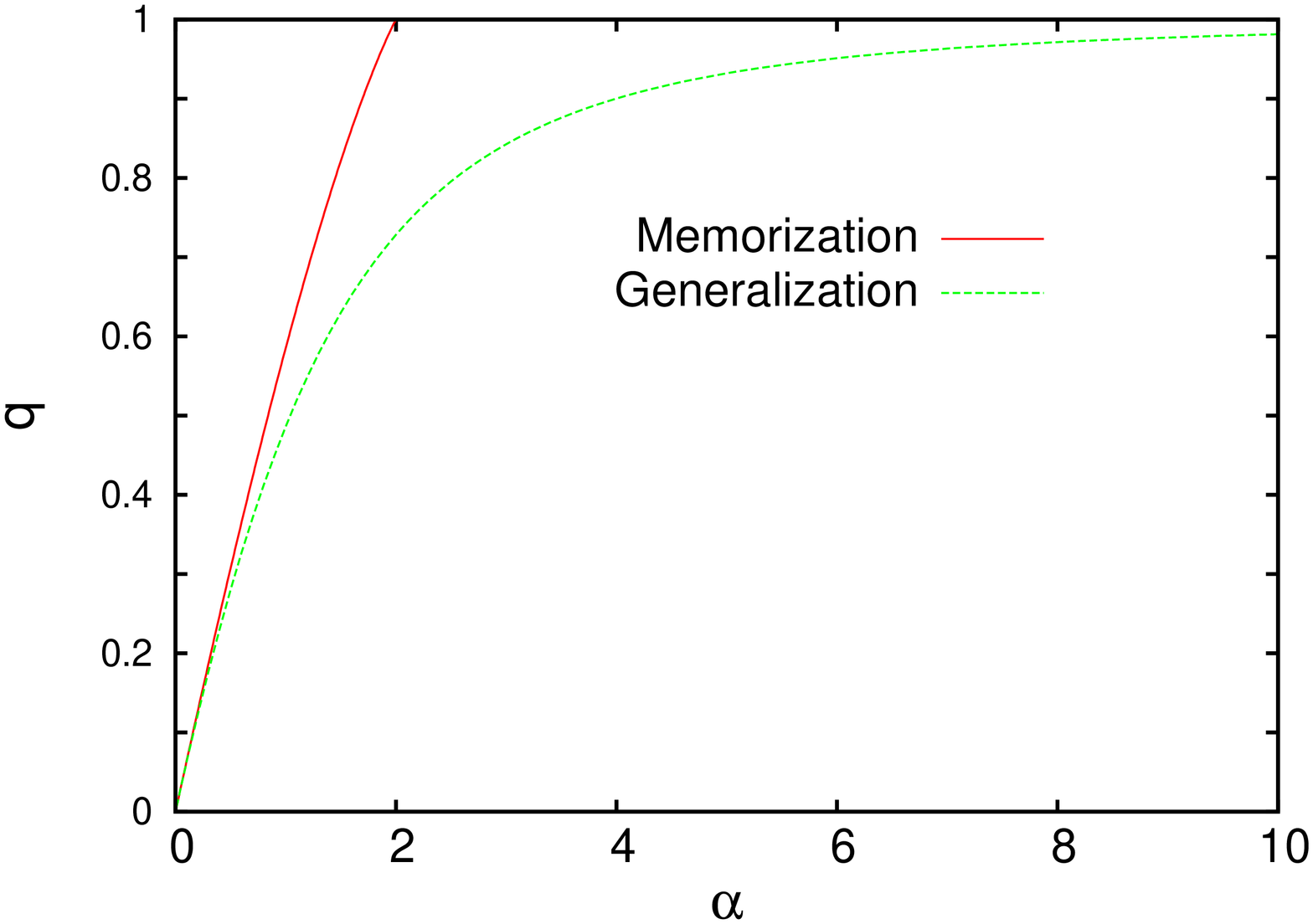}
\caption{The eigenvalues product (left) evaluated in the replica
  symmetric solution for the generalization and memorization non
  diluted. The corresponding self-overlap $q$ of the student (right).}
\label{fig:avp_ginf_cero_hcero_rs}
 \end{figure}

In the memorization limit ($\gamma \to \infty$), an RS solution of
zero energy always exists for $0 < \alpha < 2$ \cite{gardner87}. For
these values of $\alpha$ the student is capable of memorizing the
experiments, whereas for $\alpha > 2$ there are no zero energy
solutions, and, solutions are not replica symmetric. In agreement with
this known behavior (see \cite{gardner87}), the product of the
eigenvalues $\delta_1 \delta_2$, evaluated in the non diluted
memorization replica symmetric solution, is negative for $\alpha <
2$. Its value grows from $-1$ at $\alpha = 0$ to $0$ at $\alpha = 2$
(Figure \ref{fig:avp_ginf_cero_hcero_rs}), an indication that the zero
energy replica symmetric solution is stable in this range.

In the generalization limit ($\gamma = 0$) the perceptron can
asymptotically learn the classification rule provided with a big
enough amount of data as shown in \cite{lagesparse09}. However, for
any finite amount of information $\alpha$, there is not just one zero-cost student, but a continuum
of them filling a bounded region of the $J$'s space (see left panel of
Fig.~\ref{fig:spheres}). An indication of the {\em size} of the
solution space is given by the student-student overlap $q=\langle J^a
J^b\rangle$. As in the memorization case (for $\alpha<2$), the
solution space is connected, and, consistently in this case the
product $\delta_1 \delta_2$ remains always negative (see
Fig. \ref{fig:avp_ginf_cero_hcero_rs}), which means that the replica
symmetric solution is stable for every $\alpha$ as in
\cite{hungaros}. As $\alpha \to \infty$ the solutions space shrinks
around the correct value $\vecJcero$, and $q \to 1$ (see
Fig. \ref{fig:avp_ginf_cero_hcero_rs}).

As expected, as little information is given to the student (low values
of $\alpha$), there is little difference between the memorization and
the generalization. Figure \ref{fig:avp_ginf_cero_hcero_rs} shows that
also the stability, given by the product $\delta_1 \delta_2$ for the
non diluted generalization and memorization coincides for small
$\alpha$ and have the same limit for $\alpha \rightarrow 0^+$.

\subsection{Stability of diluted perceptron $h=\infty$}

The dilution field select those solutions with the lowest
values of the norm. In particular, $L_0$ and $L_1$ norms are known to
force the sparsity of the solutions \cite{LASSO}, pushing a fraction
of the students components to be zero. Seeking for sparse solutions
can enhance the efficiency in the use of available information
\cite{LASSO,mezard_compress,kabashima,kabashima2}, in particular when
the teacher is actually sparse. In the following we will consider a
$95\%$ sparse teacher, {\em i.e.} $\neffcero=0.05$ in
Eq.~(\ref{eq:rhoJcero}).

The behavior of large dilution field $h \rightarrow \infty$ limit
depends on the norm used and so does the learning behavior of the
perceptron \cite{lagesparse09}. We restrict ourselves to the study of
the space of $E=0$ solutions: to do so we take the limit $T\to 0$
first, and then the limit $h\to\infty$. Again in \cite{lagesparse09}
it was shown that, in the replica symmetric case, $h\to\infty$
pushes $q\to 1$ at any $\alpha$, concentrating the Gibbs measure over
a single point: the most diluted zero energy $\vecJ$. However, the
question remains as whether the replica symmetric ansatz is still
valid or the strong dilution field fractures the Gibbs measures into
unconnected components.

\begin{figure}[!htb]
             \includegraphics[scale=0.35, angle=270]{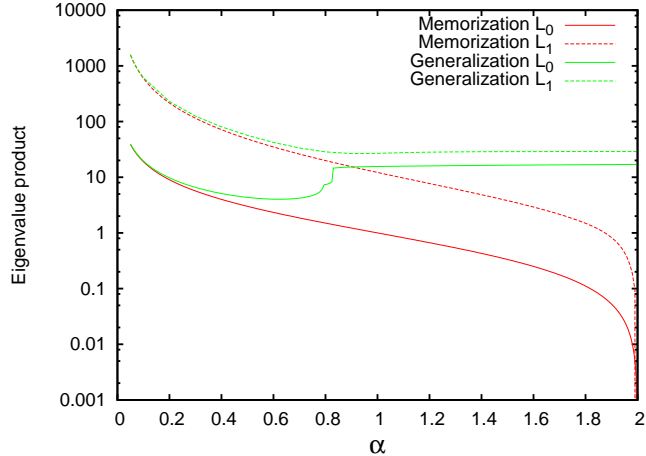}
\caption{The eigenvalues product evaluated in the replica symmetric solution for the $L_0$ and the $L_1$ norms in generalization and memorization.}
\label{fig:avp_gcero_inf_hinf_rs}
 \end{figure}

In figure \ref{fig:avp_gcero_inf_hinf_rs} we show the eigenvalues
product for the $L_0$ and $L_1$ dilutions in the $h\to\infty$ limit
for both, the memorization and the generalization cases. At variance
with the non diluted case, the $h=\infty$ one is always unstable.

\section{Phase diagram}
\label{sec:phase}

\begin{figure}[!htb]
             \includegraphics[scale=0.3, angle=270]{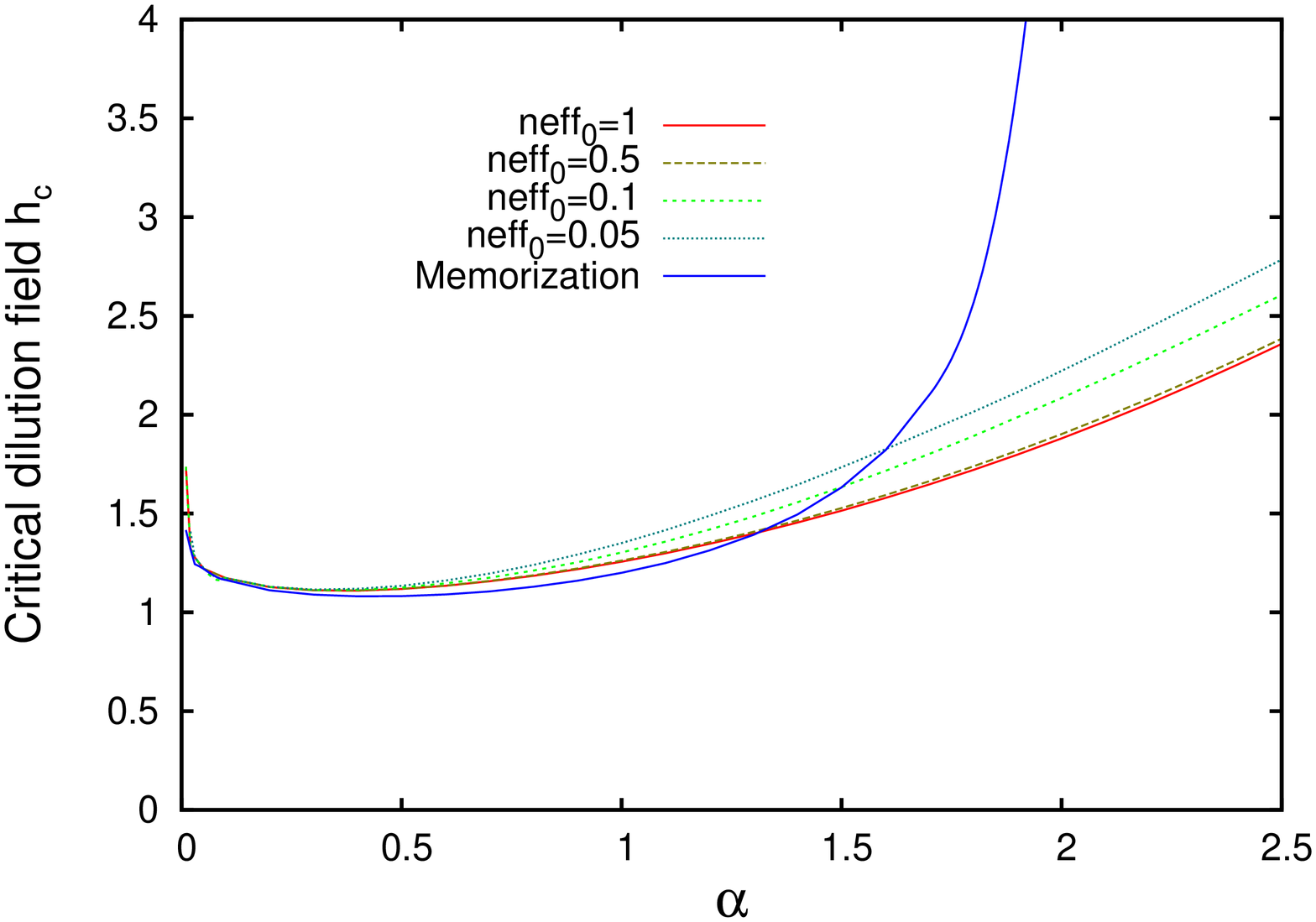}
             \includegraphics[scale=0.3, angle=270]{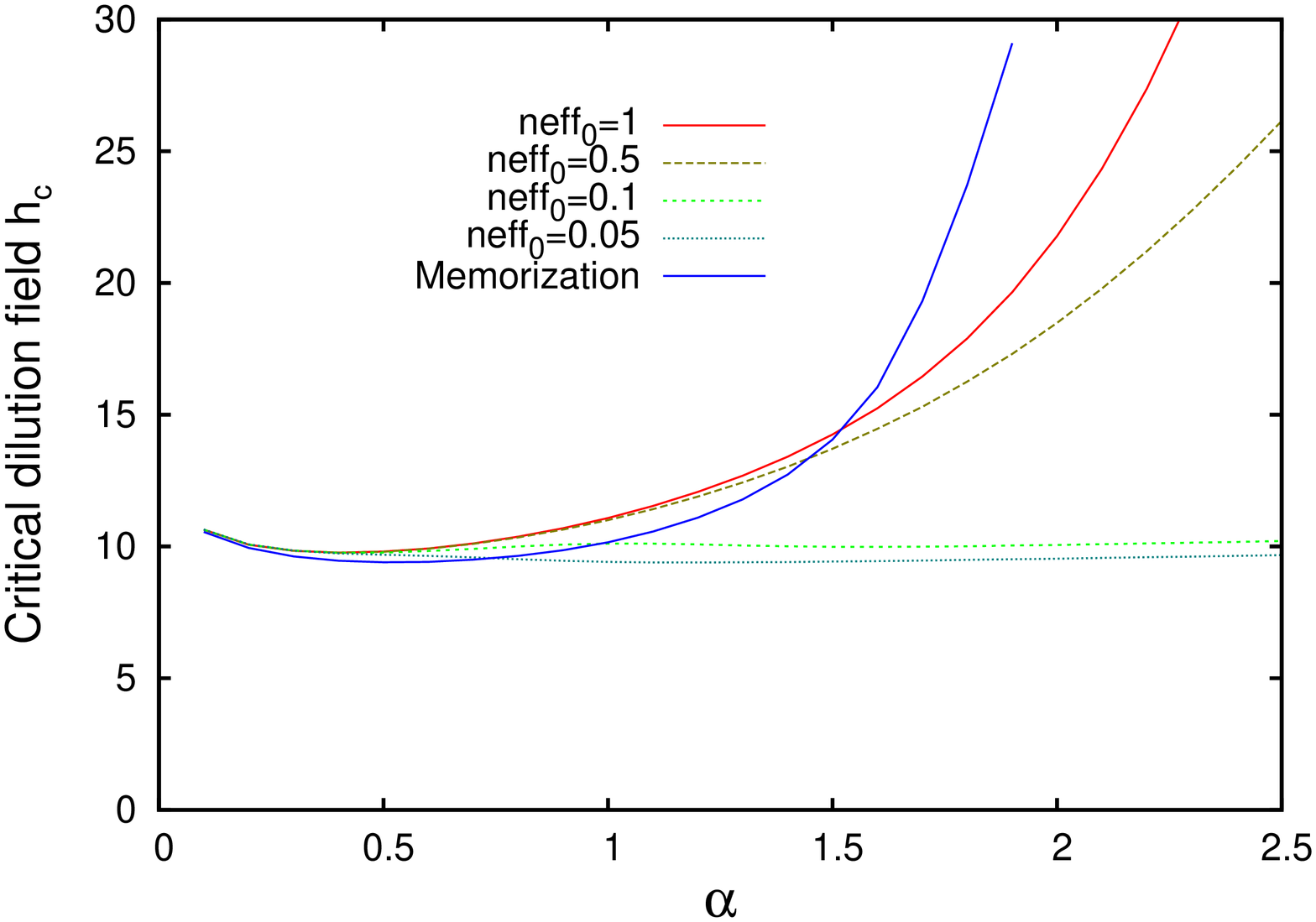}
	\caption{The phase diagram for the $L_1$(left) and the $p =
          0.1$(right) norms with severals teacher's dilution
          $n_{eff}^0$. The replica symmetric solution is stable for
          values of $\alpha$ and $h$ below the lines.}
\label{fig:df}
 \end{figure}

In the region of $\alpha$ where perfect memorization/generalization is
possible, the replica symmetric ansatz is stable in absence of
dilution and unstable in presence of a very strong dilution field
$h\to \infty$. Therefore there should exist a critical value for the
dilution field $h_c(\alpha)$ separating the RS from the RSB phase. In
figure \ref{fig:df} we show $h_c(\alpha)$ for the $L_1$ and the $p =
0.1$ norms applied to models with different teachers dilutions.

The similarity of the stability curves near $\alpha = 0$ is because
with such a few information there is not big difference between
memorizing and generalizing. The curves separate around $\alpha
\gtrsim 0.4$. The value of the critical dilution field increases with
$\alpha$. This can be rationalized as follows: more information
implies a reduction in the size of the $E=0$ solution space, and
therefore different solutions are closer, so a higher dilution field
is needed to clusterize the set of zero energy students. In
generalization, when $\alpha \rightarrow \infty$ the solution space
is composed by an only vector which is precisely the teacher and the
replica-replica overlap $q$ achieves its maximum value $1$,
consequently the critical curves go to infinity too. For the
memorization case, the value $q = 1$ is achieved for $\alpha = 2$,
that's why the critical curve diverges here.

Drawing the critical lines for the $L_0$ norm is not obvious because setting $p = 0$ in the equations, before taking the strong dilution limit $h \rightarrow \infty$, eliminates the effect of any finite dilution field, so there is no way to find $h_c$ (see \cite{lagesparse09}). What we have done instead is to study the critical lines for a value of the norm exponent $p$ small enough. The results using $p=0.1$ are shown in the right panel of Fig. \ref{fig:df}. Again, around  $\alpha = 0$ both, memorization and generalization, behave similarly. The memorization critical line also diverges when $\alpha \rightarrow 2$ while the generalization ones increases with $\alpha$ as for the $L_1$ norm. In both cases, $L_1$ and $L_{0.1}$, the critical field $h_c(\alpha)$ depends on the sparsity $\neffcero$ of the teacher (eq. (\ref{eq:rhoJcero})).

%
%

\section{Conclusions}
\label{sec:conclusions}

We have performed a full stability analysis of the replica symmetric solution for the non diluted and diluted generalization and memorization problems for a learning perceptron. Imposing a dilution should improve the results obtained in real algorithms, but comes with a price. We showed that even in the satisfiable phase (where zero cost solutions exists) an infinite dilution breaks the symmetry of the replicas, whereas no dilution at all keep the replica symmetric solution stable. The breaking of the symmetry is usually connected with convergence problems in  algorithms like belief propagation. Yet, it is always possible to find a dilution field weak enough for the replica symmetric solution to be stable, provided it exists. The critical dilution field $h_c(\alpha)$ depends on the actual sparseness of the teacher.

\begin{figure}[!htb]
             \includegraphics[scale=0.9, angle=0]{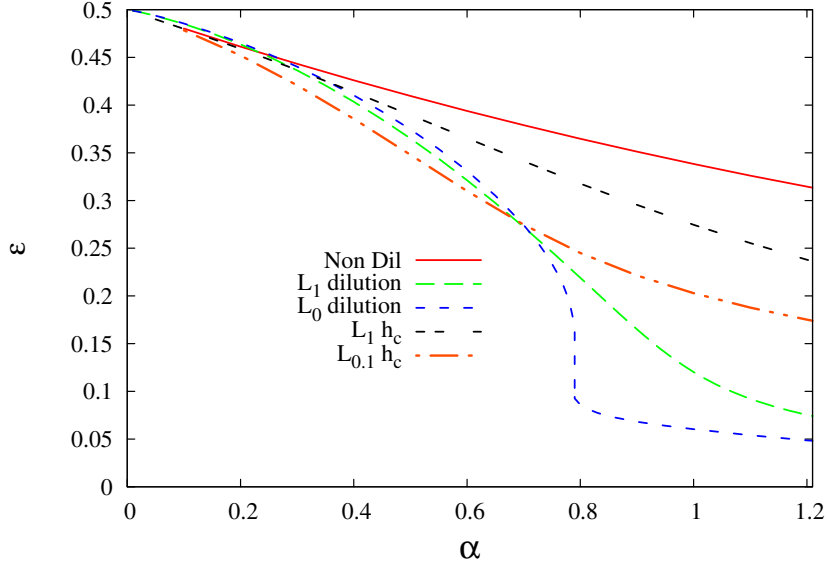}
	\caption{First three lines (in caption order) are the generalization error obtained without dilution and with strong $h\to\infty$ dilution for the $L_1$ and $L_0$ norms (results taken from \cite{lagesparse09}). The last two lines correspond to the generalization error obtained with $h=h_c(\alpha)$ in the $L_1$ and $L_{0.1}$ cases.}
\label{fig:generror}
 \end{figure}

Figure \ref{fig:generror}, partially taken from \cite{lagesparse09}, shows the generalization error achieved in average case by a perceptron without a dilution filed, and with a very strong ($h\to \infty$) dilution field. The teacher used is quite sparse ($\neffcero = 5\%$), and therefore the learning process is enhanced by the use of dilution. Restricting ourselves to the replica symmetric space, the $h\to\infty$ limit is not achievable, and $h_c(\alpha)$ is the best we can do. The gain in accuracy (lower error) obtained with an $h_c$-diluted perceptron is not as impressive as the $h\to\infty$ case, but still improves the results without dilution.

The present work is useful as it defines the phase space of the replica symmetric solution in the  information-dilution ($\alpha-h$) plane. The actual relation of this predictions to the behavior of algorithms remains as an open project for the future. Also left for the future is the use of 1RSB (or higher) parameters to make average predictions in the RSB phase.

\begin{acknowledgments}
The authors would like to thank Dr. Martin Weigt, of the Statistical Genomics and Biological Physics group in the Universit\'e Pierre et Marie Curie, for useful comments.
\end{acknowledgments}

\appendix

\section{Details of the stability calculation}
\label{app:details}

The variational free energy in terms of the all $Q_{a,b}$ and their Fourier counterparts is (see \cite{lagesparse09}), after taking the $N \rightarrow \infty$, given by
\begin{eqnarray*}
\B n \overline{f_n} &=&  - \imath \sum_{a \leq b} Q_{ab} \hatQ_{ab}-\\
&& -\log \int \prod_{a=0}^{n} \ud J^a \intrho e^{-h \sum_a \norm{J^a}^p - \imath \sum_{a \leq b} \hatQ_{ab} J^a J^b}-\\
&& - \alpha \log \int \DN{\eta} \prod_{a=0}^{n} \frac{\ud x^a \ud \hat{x^a}}{ \sqrt{2 \pi}} e^{-\beta \sum_{a=1}^n \theta(-x^a (x^0+\eta)) + \imath \sum_{a=1}^n x^a \hat{x^a} - \frac{1}{2} \sum_{a,b}^n Q_{ab} \hat{x^a} \hat{x^b}}
\end{eqnarray*}

Deriving respect to its $(n+1)(n+2)$ parameters $Q_{ab}$ and $\hatQ_{ab}$, one can construct the second derivatives matrix or Hessian of $\B n \overline{f_n}$ .The sign of the eigenvalues of this matrix evaluated in an extremal point will give us all the information about its stability.

The free energy second derivatives are
\begin{eqnarray*}
\frac{\partial^2(\B n \overline{f_n})}{\partial Q_{ab} \partial Q_{cd}} &=& \alpha (1 - \frac{\delta_{ab}}{2}) (1 - \frac{\delta_{cd}}{2}) (\langle\hat{x^a} \hat{x^b}\rangle \langle\hat{x^c} \hat{x^d}\rangle - \langle \hat{x^a} \hat{x^b} \hat{x^c} \hat{x^d} \rangle)\\
\frac{\partial^2(\B n \overline{f_n}) }{\partial (-\imath \hatQ_{ab}) \partial (-\imath \hatQ_{cd})} &=& \langle J^a J^b J^c J^d \rangle - \langle J^a J^b \rangle \langle J^c J^d \rangle \\
\frac{\partial^2(\B n \overline{f_n}) }{\partial Q_{ab} \partial (-\imath \hatQ_{cd})} &=& \delta_{ab,cd}
\end{eqnarray*}
where
\begin{eqnarray*}
\langle g(\hat{x^a}, \hat{x^b},...) \rangle&=&\frac{\int \DN{\eta} \prod_{a=0}^{n}  \frac{\ud x^a \ud \hat{x^a}}{ \sqrt{2pi}} g(\hat{x^a}, \hat{x^b},...) e^{-\beta \sum_{a=1}^n \theta(-x^a (x^0+\eta)) + \imath \sum_{a=1}^n x^a \hat{x^a} - \frac{1}{2} \sum_{a,b}^n Q_{ab} \hat{x^a} \hat{x^b}}}{\int \DN{\eta} \prod_{a=0}^{n} \frac{\ud x^a \ud \hat{x^a}}{ \sqrt{2pi}} e^{-\beta \sum_{a=1}^n \theta(-x^a (x^0+\eta)) + \imath \sum_{a=1}^n x^a \hat{x^a} - \frac{1}{2} \sum_{a,b}^n Q_{ab} \hat{x^a} \hat{x^b}}}
\end{eqnarray*}

\begin{eqnarray*}
\langle g(J^a, J^b,...) \rangle&=&\frac{\int \prod_{a=0}^{n} \ud J^a \intrho g(J^a, J^b,...)e^{-h \sum_a \norm{J^a}^p - \imath \sum_{a \leq b} \hatQ_{ab} J^a J^b}}{\int \prod_{a=0}^{n} \ud J^a \intrho e^{-h \sum_a \norm{J^a}^p - \imath \sum_{a \leq b} \hatQ_{ab} J^a J^b}}
\end{eqnarray*}

The structure of the Hessian matrix is:
{\small\begin{equation}
\|\partial^2(\B n \overline{f_n})\| = \left(\begin{array}{ccccccc}
\|\frac{\partial^2(\B n \overline{f_n}) }{\partial Q_{ab} \partial Q_{cd}}\|&&&\|\frac{\partial^2(\B n \overline{f_n}) }{\partial Q_{ab} \partial (-\imath \hatQ_{cd})}\|\\
\|\frac{\partial^2(\B n \overline{f_n}) }{\partial Q_{ab} \partial (-\imath \hatQ_{cd})}\|&&&\|\frac{\partial^2(\B n \overline{f_n}) }{\partial (-\imath \hatQ_{ab}) \partial (-\imath \hatQ_{cd})}\|
\end{array}\right) \label{eq:hessian1}
       \end{equation}

The eigenvector space of the Hessian matrix can be divided in two \cite{gardner87,statistical_mechanics_of_learning,almeida}; one subspace corresponding to instabilities inside the replica symmetric space (and whose eigenvalues signs are those who determines among of all the possible replica symmetric solutions the one who actually minimizes the free energy), and the other corresponding to instabilities that takes our extremal point outside the replica symmetric space. The first of these subspaces has no interest to our stability analysis, because solving the replica symmetric fixed point equations is equivalent to searching for the stable extremal replica symmetric point, so, we are going to focus on the eigenvalues of the eigenvectors belonging to the second subspace. A detailed study of the Hessian matrix shows that it has just two non replica symmetric values, $\delta_1$ and $\delta_2$. The product $\delta_1 \delta_2$ can be expressed as in eq. (\ref{eq:eigenvalue_product}) in terms of the eigenvalues of the inner matrices.

\bibliography{biblio}
\end{document}